# Poster Abstract: Combining edge and cloud computing for mobility analytics

Ikechukwu Maduako
imaduako@unb.ca

Hung Cao
hcao3@unb.ca

Lilian Hernandez
lhernand@unb.ca

Monica Wachowicz
monicaw@unb.ca

People in Motion Lab, University of New Brunswick
15 Dineen Drive, Fredericton, NB. E3B 5A3 Canada

## I. INTRODUCTION

Mobility analytics using data generated from the Internet of Mobile Things (IoMT) is facing many challenges which range from the ingestion of data streams coming from a vast number of fog nodes and IoMT devices to avoiding overflowing the cloud with useless massive data streams that can trigger bottlenecks [1]. Managing data flow is becoming an important part of the IoMT because it will dictate in which platform analytical tasks should run in the future. Data flows are usually a sequence of out-of-order tuples with a high data input rate, and mobility analytics requires a real-time flow of data in both directions, from the edge to the cloud, and vice-versa. Before pulling the data streams to the cloud, edge data stream processing is needed for detecting missing, broken, and duplicated tuples in addition to recognize tuples whose arrival time is out of order. Analytical tasks such as data filtering, data cleaning and low-level data contextualization can be executed at the edge of a network. In contrast, more complex analytical tasks such as graph processing can be deployed in the cloud, and the results of ad-hoc queries and streaming graph analytics can be pushed to the edge as needed by a user application. Graphs are efficient representations used in mobility analytics because they unify knowledge about connectivity, proximity and interaction among moving things.

This poster describes the preliminary results from our experimental prototype developed for supporting transit systems, in which edge and cloud computing are combined to process transit data streams forwarded from fog nodes into a cloud. The motivation of this research is to understand how to perform meaningfulness mobility analytics on transit feeds by combining cloud and fog computing architectures in order to improve fleet management, mass transit and remote asset monitoring [2].

## II. SYSTEM ARCHITECTURE

Our prototype system architecture is shown in Fig. 1 and consists of three-layers named as Fog Node Cluster (sensing layer), Aggregated Fog Node (access layer) and Cloud (core layer). The fog node cluster is a group of mobile fog nodes that pull data from sensors deployed in a transit vehicle. Each mobile fog node is designed to be installed inside a transit vehicle belonging to a transit fleet. The Cisco 829 GW-LTE-NA-AK9 was selected for the experiment because is designed for harsh environments including shock, humidity, and wide temperature range. The transit data streams contain information including bus route identifier, bus route number, vehicle identifier, GPS coordinates and timestamp that are generated at different time granularities, including every 5s, every 30min, or every day (Fig 2.)

We selected an aggregated fog node for running the Edge Fog Fabric (EFF) platform in order to pass the data streams from the mobile fog node cluster to the cloud, and vice-versa [3]. The EFF manages and ensures that there is an appropriate flow of unbounded tuples in both directions, and perform analytical tasks such as filtering, cleaning and contextualization. The EFF is basically composed by a system administrator, dataflow editor and engine, system monitor, message broker, links, and IoT database (i.e. ParStream) [4].

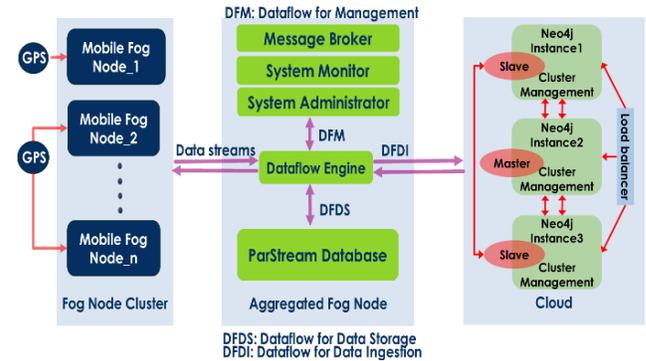

**Figure 1**. Overall prototype architecture.

The cloud is the core layer where the Neo4j database is situated. It is the place where mobility analytics is conducted based on the compressed and post-processed data coming from the EFF. The mobility analytics is carried out using an integrated Apache Spark GraphX engine and Neo4J.

Fig. 2 illustrates the life cycle that takes place every time the tuples flow through the system, from the mobile fog nodes to the cloud. It consists of six steps: the data transportation, the data processing, the data leverage, the data control, the data acquisition, and the data storage. Every tuple is sent (data transportation) and received by the destination mobile fog nodes (data acquisition). After that, in order to process the tuple (data processing) and perform mobility analytics, the EFF controls the sets of tuples (data control) by retaining (data storage) and retrieving (data leverage) them continuously. All the steps related in this life cycle of our prototype are operated through the ParStream [3].

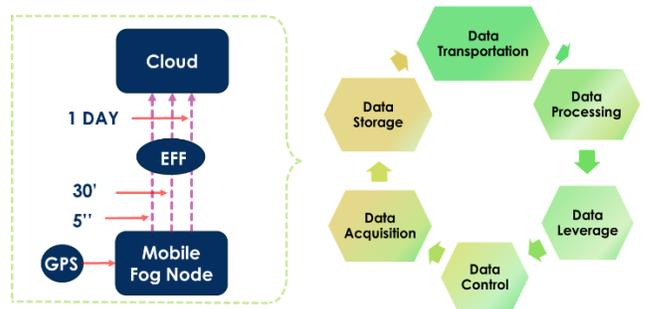

**Figure 2**. Life cycle for IoMT of the experimental system.

## III. MOBILITY ANALYTICS

The mobility analytical workflow consists of: (1) data cleaning tasks deployed at the mobile fog nodes (sensing layer); (2) data contextualization tasks performed at the aggregated fog node (access layer), and (3) graph query running in the cloud (core layer). Data cleaning is always necessary in order to remove errors and inconsistencies from the tuples. The data cleaning task is implemented using a Python script algorithm for processing five automated steps to handle *(1) missing tuples, (2) duplicated tuples, (3) missing attribute values, (4) redundant attributes, and (5) wrong attribute values* [5].

The data contextualization task enriches the tuples from the previous data cleaning task using higher level concepts accordingly to a particular mobility context. In the current prototype, the geographical coordinates (x, y) and the timestamp t of each tuple are used for this contextualization. First, an empirical distance value of 15m is designated to compute stops and moves. Second, the Euclidean distance between two consecutive points (i.e. tuples) is computed. If the distance between them is smaller than 15m, a new attribute containing the value *"stop"* is added to the second tuple. In contrast, if the distance is higher than 15m, the *"move"* attribute value is added to the second tuple.

Finally, time-varying graph queries using graph metrics such as shortest path, degree and page rank centrality are run in a Neo4j database in the cloud, which includes a master node and slave nodes that were deployed using the Compute Canada West Cloud [6]. The query outputs are essentially a time-series of static graph snapshots based on a time tree, which is effective for online mobility analytics of large temporal graphs where handling speed and complexity is at the most importance.

## IV. PRELIMINARY RESULTS

The transit feeds generated by CODIAC transit network for the Greater Moncton area was used for the implementation of our prototype system. The transit fleet consists of 642 bus stations belonging to 30 bus routes operating from Monday to Saturday, some of which also providing evening and Sunday services.

Fig. 3 shows the results after running a shortest-path query and retrieving the shortest bus trips at different peak hours on June 8[th] 2016. In this figure, red nodes represent stops which might occur because of a traffic jam, accident, collecting passengers at a bus station, or a traffic light at one street intersection. Green nodes represent moves that occur because a bus is moving on a street or passing by a bus station because there are no passengers to drop off or get on.

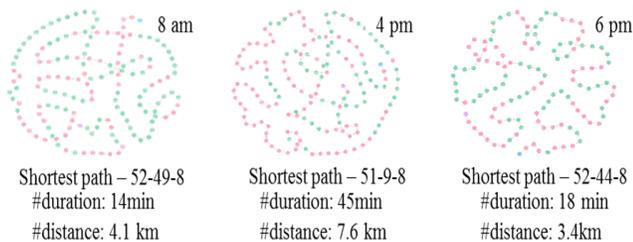

**Figure 3**. The trip dynamics of the same bus route over time.

Fig. 4 shows the degree and Page Rank query results for retrieving information about the dynamics of the bus stations of a bus route. The largest number of stops (red nodes) clustered around a bus station (grey nodes) indicates where stopovers have occurred for collecting and dropping passengers. The page rank score results show the busiest stations as being the ones located at the Plaza, Champlain Street and Main Street. The moves (green nodes) shows the bus stations where buses passed by a bus station because there were no passengers to drop off or get on.

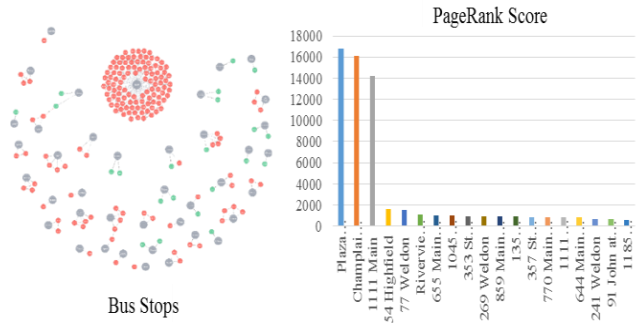

**Figure 4**. The trip dynamics of bus stations.

## V. FUTURE RESEARCHWORK

This paper describes the potential of mobility analytics to be performed over data streams based on the concept of light-weight data processing at the edge, and heavy graph processing in the cloud. The preliminary results are positive in paving the way for developing edge-cloud system architectures that can support data flows for specific purposes, e.g. a dataflow for processing, a dataflow for data storage, and a dataflow for data ingestion. Each dataflow must be designed taking into account a mobility analytics task. We have implemented only one single dataflow (i.e. from mobile fog nodes to the cloud), therefore future research work will be focused on both directions. We have also implemented the EFF platform at the access layer, but we expect to deploy it at the other layers as well. The research challenge will be to design a coherent mobility analytical workflow which can handle the scalability, speed and complexity issues of data streams.

### ACKNOWLEDGMENTS

This research was fully supported by the NSERC/CISCO Industrial Research Chair in Real-time Mobility Analytics. The authors are grateful to CODIAC Transit for providing the data streams used in this study, and Compute Canada for hosting one virtual machine that was used for the implementation of the cloud layer. Finally, we would like to thank Opio for their support in the mobile fog node configuration.

### REFERENCES

[1] Gama, J., and Gaber, M.M. eds., 2007. Learning from Data Streams: Processing Techniques in Sensor Networks. *Springer Science & Business Media*. 25–50.
[2] Cao, H., Wachowicz, M., and Cha, S., 2017. Developing an edge analytics platform for analyzing real-time transit data streams. *arXiv preprint arXiv:1705.08449*.
[3] Cisco white paper, 2016. The Cisco edge analytics fabric system: A new approach for enabling hyper distributed implementations. *Cisco public*, 1–22, in press.
[4] Cisco, 2017. The Cisco Parstream manual. *Cisco public*, Version 4.4.3, 16–33.
[5] Cao, H. and Wachowicz, M., 2017. The design of a streaming analytical workflow for processing massive transit feeds. *arXiv preprint arXiv: 1706.04722*.
[6] Cha, S., Ruiz, M.P., Wachowicz, M., Tran, L.H., Cao, H. and Maduako, I., 2016, December. The role of an IoT platform in the design of real-time recommender systems. In *Internet of Things (WF-IoT), 2016 IEEE 3rd World Forum on*, 448-453. IEEE.